\documentclass[conference]{IEEEtran}
\usepackage[numbers]{natbib}
\usepackage{graphicx}
\usepackage{amsmath}
\usepackage{verbatim}
\usepackage{natbib}
\usepackage{hyperref}
\usepackage{caption}
\usepackage{subcaption}
\usepackage{multirow}

\ifCLASSINFOpdf
\else
\fi
\begin{document}
%
\title{Multimedia Distribution Process Tracking for Android and iOS}

\author{\IEEEauthorblockN{Yu-Min Jeon}
\IEEEauthorblockA{9th Best of the Best program, KITRI\\
Seoul, South Korea \\
Email: men1999@naver.com}
\and
\IEEEauthorblockN{Won-Mu Heo}
\IEEEauthorblockA{School of Computer Software,\\
Daegu Catholic University, \\ South Korea\\
Email: hwm1308@gmail.com}
\and
\IEEEauthorblockN{Jong-Min Kim}
\IEEEauthorblockA{Department of Computer \\ Science and Engineering,\\
Korea University, South Korea\\
Email: dakuo@korea.ac.kr}
\and
\IEEEauthorblockN{Kyounggon Kim}
\IEEEauthorblockA{Center of Excellence in Cybercrime and Digital Forensics, \\
Naif Arab University for Security Sciences, \\ Riyadh, Kingdom of Saudi Arabia \\ 
Email: kkim@nauss.edu.sa}}


%


\maketitle

\begin{abstract}
The crime of illegally filming and distributing images or videos worldwide is increasing day by day. 
With the increasing penetration rate of smartphones, there has been a rise in crimes involving secretly taking pictures of people's bodies and distributing them through messengers.
However, little research has been done on these related issue.
The crime of distributing media using the world's popular messengers, WhatsApp and Telegram, is continuously increasing. 
It is also common to see criminals distributing illegal footage through various messengers to avoid being caught in the investigation network. 
As these crimes increase, there will continue to be a need for professional investigative personnel, and the time required for criminal investigations will continue to increase.
In this paper, we propose a multimedia forensic method for tracking footprints by checking the media information that changes when images and videos shot with a smartphone are transmitted through instant messengers. 
We have selected 11 of the world's most popular instant messengers and two secure messengers. 
In addition, we selected the most widely used Android and iOS operating systems for smartphones. 
Through this study, we were able to confirm that it is possible to trace footprints related to the distribution of instant messengers by analyzing transmitted images and videos. 
Thus, it was possible to determine which messengers were used to distribute the video when it was transmitted through multiple messengers. 
\end{abstract}


%
\IEEEpeerreviewmaketitle

\section{Introduction}
As the penetration rate of smartphones has increased, crimes such as simulation and fraud targeting on Online Social Networks (OSNs) using smartphones have also evolved.
Patel et al. conducted research indicating that social media and messaging platforms are currently being used as tools for cybercrime. Therefore, further research on OSNs and messaging platform is considered important \cite{patel2017theoretical}.

Existing research on criminal investigation focuses on artifact extraction from specific apps, text mining, and image quality analysis. 
While existing studies allow for the retrieval of records transmitted through artifacts in a specific app, there is insufficient research on identifying the source and path of media, such as images and videos, when they are not recorded on the device. 
This gap shows that current methods may be inadequate for responding to the distribution of illegally filmed footage, which is a type of crime. 
The spread of pornography is particularly concerning, as it can be easily shared in messenger chat rooms.
According to the \cite{Multiple2020} report, a recent study found that more than 42\% of adults use two or more social media platforms in their daily lives.  
As a result, there is a high likelihood that such content will be shared again on other applications. 

Given that it is impossible for investigators to manually check each image or video, it is necessary to identify the source and route of the media through media analysis. 
In this study, we examined a method for investigating the source of an image received through a messenger on a mobile device. 
This information is crucial for investigators to trace the origin of media found on a suspect's mobile device. 
This is because the media found on the device was either received from someone else or obtained from elsewhere, unless it was filmed using the device itself. 

In the initial stages of this study, we considered four questions:
i) What media artifacts are left behind in a specific app?; 
ii) Do artifacts differ for each smartphone?;
iii) Is it possible to identify the platform through which media has been transmitted?;
iv) How could the alteration of media evidence through transfers affect criminal investigations?

To the best of our knowledge, there is no other study that analyzes the media changes that occur when a specific messenger is passed through multiple messenger platforms. The contributions of this paper are as follows:
\begin{enumerate}
\item First, we have classified encoded media artifacts for eleven of the most popular messengers and two secure messengers.
\item Second, we analyzed the changes in status when media encoded once in a messenger passes through another messenger.
\item Third, we suggested investigative methods based on the analysis results.
\end{enumerate} 

The remainder of this paper is structured as follows: Section 2 introduces existing research, Section 3 presents our research methodology, Section 4 evaluates the study results, and finally, Section 5 summarizes the conclusions and discusses future directions. 

\section{Related work}
Existing Instant Messenger (IM)-related studies have mainly focused on artifact analysis, image quality analysis, and text mining research that analyzes text in messengers \cite{son2022forensic}. 
Over the past 10 years, with the increasing popularity of social networking services (SNS) and messaging apps, many studies on artifact analysis have been conducted \cite{campana2021mydigitalfootprint}. 
These studies have successfully identified artifacts for specific apps, such as an account information, text messages, photos, calls, and chats, especially on Android compared to iOS.

\cite{alisabeth2020forensic} showed that user configurations containing information related to user accounts, follower accounts, and close friend accounts were identified for Instagram. 
\cite{chang2020evidence} showed evidence collected, including text, photo, video transmission, and calls from Facebook Messenger. 
Similar studies have been conducted in other applications such as Discord (\cite{motylinski2020digital}), LINE (\cite{chang2019line, riadi2018evidence, chang2018forensic, riadi2019live}), WhatsApp (\cite{ghannam2018forensic, orr2018whatsapp}), Telegram (\cite{anglano2017forensic, satrya2016digital, hintea2018forensic, gregorio2017forensic, gregorio2018forensic}), KakaoTalk (\cite{choi2017forensic, yoon2016forensic}), Facebook (\cite{agrawal2019digital}), Skype (\cite{al2013skype, nicoletti2019forensic}), Signal (\cite{judge2018mobile, krishnapriya2021forensic}), WeChat (\cite{wu2017forensic, zhang2016forensic}), QQ (\cite{li2017flexibility, gao2010memory}), ChatON (\cite{iqbal2013forensic}), Kik (\cite{adebayo2017forensic, ovens2016forensic}), and Wickr (\cite{mehrotra2013forensic, kim2021forensic, azhar2017forensic}).

As the studies mentioned above show, artifact analysis for a specific app is currently in progress. 
However, since there are many different messaging apps and most people use several types of messengers, further research is needed.
\cite{anwar2021image} addressed that the issue of message transmission of PNG images, which is secured using end-to-end encryption and compressed according to predefined rules. 
This study analyzed PNG image compression and alpha channel by comparing and analyzing PNG images that were sent through WhatsApp and PNG images before transmission. 
However, it did not deal with rhetorical meaning.

On the other hand, \cite{marfianto2018whatsapp} built a web system for WhatsApp that uses text mining methods to provide information about instructions related to the intervention of others. 
The method used for message identification is to calculate the weight of each word for the transmitted document using TF-IDF and cosine similarity.

\section{Methodology}
Through the analysis of media disseminated via instant messengers, we proposed a methodology to identify footprints related to the characteristic ``the media is distributed from a certain messenger''.
We conducted scenarios assuming that the media shot on the smartphone was distributed through the application. 
In addition, there are several processes involved in dissemination of media, such as cloud, drive, USB, and software, and the specific dissemination process is unknown. 
However, by checking the media artifacts left by a specific messenger, it is possible to track the distributed messenger through footprinting. 

We used MediaInfo version 20.09 as a media analysis tool. 
iOS supports both high-efficiency formats and high-compatibility formats. 
However, since the high-efficiency format requires an OS version of 11 or higher and an iPhone 7 or higher, we conducted the experiment by selecting a high-compatibility format to further generalize the experimental results.
For Android, we experimented with two methods: the default 4:3 (12MP) 4032x3024 resolution mainly used in the new Galaxy S series and the default 16:9 (16MP) 5312x2988 resolution mainly used in the old Galaxy Note series. 
The original information on image media is shown in Table~\ref{table:table1}.

\begin{table}[h!]
\centering\small
\caption{Original image information}\label{tbl1}
\begin{tabular}{p{2cm} p{2cm} p{2cm}}
\hline
\textbf{Distribution} & iOS       & Android                                                       \\ \hline
\textbf{Extension}    & JPG       & JPG                                                           \\ \hline
\textbf{Format}       & JPEG      & JPEG                                                          \\ \hline
\textbf{Resolution}   & 4032x3024 & 
	\begin{tabular}[c]{@{}c@{}}4032x3024\\ 5312x2988\end{tabular} \\ \hline
\end{tabular}
\label{table:table1}
\end{table}

\begin{figure*}
\centering\small
	\subfloat{\includegraphics{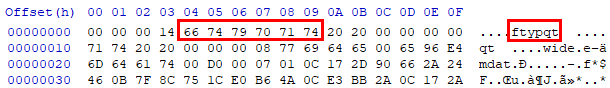}}
    \label{iOS Codec}
    \subfloat{\includegraphics{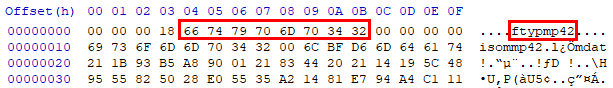}}
    \caption{Codec ID (Above iOS, Below Android)}
	\label{Android Codec}
\label{figure:figure1}
\end{figure*}

We set the camera options in iOS to 1080p HD – 30fps, and on Android to FHD 1920x1080 resolution. 
The main distinguishing features between iOS and Android are the extensions and Codec IDs. 
For extensions, iOS uses the MOV extension and Android uses the mp4 extension. 
Codec ID are presented in the form of ftypqt for iOS and ftypmp42 for Android at the File Signature using HxD, as shown in Figure~\ref{figure:figure1}.
Additionally, each video has a specific file size; for instance, a 30-second video is about 50-60 MB in size.
The video format profile can be divided into High (high quality), Main (medium quality), and Baseline (low quality). Table~\ref{table:table2} shows the original information on video media.

\begin{table}[h!]
\centering\small
\caption{Original video information}\label{tbl2}
	\begin{tabular}{p{3cm} p{2cm} p{3cm}}
		\hline
		\textbf{Distribution}	& iOS       &  Android    \\ \hline
		\textbf{Extension}      & MOV       &  mp4        \\ \hline
		\textbf{Format}         & MPEG-4    &  MPEG-4     \\ \hline
		\textbf{Format Profile} & QuickTime &  Base Media Version 2 \\ \hline
		\textbf{Codec ID}       & qt        & mp42 (isom/mp42)   \\ \hline
		\textbf{Video Format Profile} & High@L4   & High@L4       \\ \hline
		\textbf{Resolution}     & 1920x1080 & 1920x1080        \\ \hline
	\end{tabular}
\label{table:table2}
\end{table}

\section{Experiment}
We have summarized the changes made to images and videos sent from a total of 13 instant messengers for Android and iOS, which are smartphone operating systems.
In this study, KakaoTalk and Facebook Messenger were selected as the N-th messengers, and media information that was changed when media was transmitted to the rest of the messengers as the N+1st messenger was further analyzed. 

\subsection{Experiment environment}
For the media artifact analysis, we have conducted experiment on smartphones (Android, iOS), which are devices typically used for instant messengers. Table~\ref{table:table3} shows the targeted test devices and versions.

\begin{table}
\centering\small
\caption{Experimental device and version}\label{tbl3}
	\begin{tabular}{p{2cm} p{2cm} p{2cm}}
		\hline
		{\textbf{Manufacurer}} & {\textbf{Model name}} & {\textbf{OS Version}} \\ \hline
		Apple     & A1687     & 14.2          \\ \hline
		Apple     & A1905     & 14.2          \\ \hline
		SAMSUNG   & SM-N916L  & 6.0.1         \\ \hline
		SAMSUNG   & SM-G906K  & 6.0.1         \\ \hline
		SAMSUNG   & SM-G930S  & 8.0.0         \\ \hline
	\end{tabular}
\label{table:table3}
\end{table}

In this study, we analyzed the changes in media on 13 platforms, including popular messengers like WhatsApp, Facebook Messenger, and Telegram, as well as some secure and lesser-known platforms.
The mobile messenger applications used in the experiment are listed in Table~\ref{table:table4}.

\begin{table}[h!]
\centering\small
\caption{Mobile messenger application information}\label{tbl4}
\begin{tabular}{p{2.7cm} p{2cm} p{2.5cm}}
\hline
	\textbf{Application name}  & \textbf{Version (iOS)} & \textbf{Version (Android)} \\ \hline
	KakaoTalk       & 9.1.2      & 9.1.3                      \\ \hline
	Facebook Messenger & 291.2                  & 291.2.0.22.114             \\ \hline
	Facebook       & 297.0       & 297.0.0.36.116             \\ \hline
	Instagram      & 167.0       & 167.1.0.25.120             \\ \hline
	WhatsApp       & 2.20.121    & 2.20.205.16                \\ \hline
	WeChat         & 7.0.18      & 7.0.17                     \\ \hline
	Telegram       & 7.2.1       & 7.2.1                      \\ \hline
	Skype          & 8.66.76     & 8.66.0.76                  \\ \hline
	Discord        & 49.0        & 49.16                      \\ \hline
	NateOn         & 4.0.5       & 4.0.8                      \\ \hline
	LINE           & 10.19.0     & 10.20.1                    \\ \hline
	Signal         & 3.22.0      & 4.78.5                     \\ \hline
	Wickr Me       & 5.66.10     & 5.66.8                     \\ \hline
\end{tabular}
\label{table:table4}
\end{table}

\begin{table}[h!]
\centering\small
\caption{Software and web browser information}\label{tbl5}
\begin{tabular}{p{2.7cm} p{2.3cm} p{2.1cm}}
\hline
\textbf{Application name}    & \textbf{Version (Software)} & \textbf{Browser (Web)} \\ \hline
KakaoTalk      & 3.1.9.2626     &              \\ \hline
WhatsApp       & 2.2047.11      &              \\ \hline
WeChat         & 3.0.0.57       &	           \\ \hline
Telegram       & 2.4.7          &       	   \\ \hline
Skype          & 15.66.77.0     &      		   \\ \hline
Discord        & 0.0.308        &       	   \\ \hline
NateOn         & 7.0.3.1        &              \\ \hline
LINE           & 6.4.0.2388     &      	       \\ \hline
Signal         & 1.37.2         &       	   \\ \hline
Wickr Me       & 5.65.4         &              \\ \hline
Facebook Messenger &            & Chrome       \\ \hline
Facebook       &  	  	        & Chrome       \\ \hline
Instagram      &                & Chrome       \\ \hline
\end{tabular}
\label{table:table5}
\end{table} 

Facebook Messenger, Facebook, and Instagram are primarily used in a PC environment through web browsers rather than software.
As a result, we downloaded media from 10 messengers using software and from 3 messengers through a web browser.
The software and web browser information used in the study are presented in Table~\ref{table:table5}.

\begin{table*}
\centering\small
\caption{Change image in instant messengers}\label{tbl6}
	\begin{tabular}{|p{2.5cm}|p{2cm}|p{1.5cm}|p{1.5cm}|p{1.5cm}|p{1.5cm}|p{1.5cm}|p{1.5cm}|}
	\hline
	\multirow{2}{*}{Application name}                            & \multicolumn{1}{c|}{\multirow{2}{*}{Image Option}} & \multicolumn{2}{c|}{iOS}                                        & \multicolumn{2}{l|}{Android (4032x3024)}                                                          & \multicolumn{2}{l|}{Android (5312x2988)}                                                                                                           \\ \cline{3-8} 
	& \multicolumn{1}{c|}{}                              & \multicolumn{1}{c|}{Width}     & \multicolumn{1}{c|}{Length}    & \multicolumn{1}{c|}{Width}                      & \multicolumn{1}{c|}{Length}                     & \multicolumn{1}{c|}{Width}                                                         & Length                                                        \\ \hline
	\multirow{2}{*}{KakaoTalk}                                   & \multicolumn{1}{c|}{General}                       & \multicolumn{1}{c|}{960x720}   & \multicolumn{1}{c|}{720x960}   & \multicolumn{1}{c|}{960x720}                    & \multicolumn{1}{c|}{720x960}                    & \multicolumn{1}{c|}{960x540}                                                       & 540x960                                                       \\ \cline{2-8} 
	& \multicolumn{1}{c|}{High}                          & \multicolumn{1}{c|}{1440x1080} & \multicolumn{1}{c|}{1080x1440} & \multicolumn{1}{c|}{1440x1080}                  & \multicolumn{1}{c|}{1080x1440}                  & \multicolumn{1}{c|}{1440x810}                                                      & 810x1440                                                      \\ \hline
	\begin{tabular}[c]{@{}c@{}}Facebook\\ Messenger\end{tabular} & \multicolumn{1}{c|}{Default}                       & \multicolumn{4}{c|}{Indistinguishable}                                                                                                                              & \multicolumn{1}{c|}{\begin{tabular}[c]{@{}c@{}}2048x1152\\ 3984x2241\end{tabular}} & \begin{tabular}[c]{@{}c@{}}1151x2048\\ 2240x3984\end{tabular} \\ \hline
	Facebook                                                     & \multicolumn{1}{c|}{Default}                       & \multicolumn{1}{c|}{960x720}   & \multicolumn{1}{c|}{720x960}   & \multicolumn{4}{c|}{Indistinguishable}                                                                                                                                                                                                                 \\ \hline
	Instagram                                                    & \multicolumn{7}{c|}{Indistinguishable}                                                                                                                                                                                                                                                                                                                                        \\ \hline
	WhatsApp                                                     & \multicolumn{1}{c|}{Default}                       & \multicolumn{1}{c|}{1600x1200} & \multicolumn{1}{c|}{1200x1600} & \multicolumn{1}{c|}{1600x1200}                  & \multicolumn{1}{c|}{1200x1600}                  & \multicolumn{1}{c|}{1328x747}                                                      & 747x1328                                                      \\ \hline
	WeChat                                                       & \multicolumn{1}{c|}{General}                       & \multicolumn{1}{c|}{1440x1080} & \multicolumn{1}{c|}{1080x1440} & \multicolumn{1}{c|}{1080x1440}                  & \multicolumn{1}{c|}{1080x1440}                  & \multicolumn{1}{c|}{1920x1080}                                                     & 1080x1920                                                     \\ \hline
	Telegram                                                     & \multicolumn{1}{c|}{General}                       & \multicolumn{1}{c|}{1280x960}  & \multicolumn{1}{c|}{960x1280}  & \multicolumn{1}{c|}{1280x960}                   & \multicolumn{1}{c|}{960x1280}                   & \multicolumn{1}{c|}{1280x960}                                                      & 960x1280                                                      \\ \hline
	Skype                                                        & \multicolumn{1}{c|}{Default}                       & \multicolumn{1}{c|}{2048x1536} & \multicolumn{1}{c|}{1536x2048} & \multicolumn{1}{c|}{2016x1512}                  & \multicolumn{1}{c|}{1512x2016}                  & \multicolumn{1}{c|}{1992x1121}                                                     & 1121x1992                                                     \\ \hline
	Discord                                                      & \multicolumn{7}{c|}{Indistinguishable}                                                                                                                                                                                                                                                                                                                                        \\ \hline
	\multirow{2}{*}{NateOn}                                      & \multicolumn{1}{c|}{Minimal}                       & \multicolumn{4}{c|}{\multirow{2}{*}{Indistinguishable}}                                                                                                             & \multicolumn{1}{c|}{664x374}                                                       & 374x664                                                       \\ \cline{2-2} \cline{7-8} 
	& \multicolumn{1}{c|}{Standard}                      & \multicolumn{4}{c|}{}                                                                                                                                               & \multicolumn{1}{c|}{1328x747}                                                      & 747x1328                                                      \\ \hline
	\multirow{2}{*}{LINE}                                        & \multicolumn{1}{c|}{General}                       & \multicolumn{1}{c|}{1478x1108} & \multicolumn{1}{c|}{1108x1478} & \multicolumn{1}{c|}{\multirow{2}{*}{1477x1108}} & \multicolumn{1}{c|}{\multirow{2}{*}{1108x1477}} & \multicolumn{1}{c|}{\multirow{2}{*}{1706x960}}                                     & \multirow{2}{*}{960x1706}                                     \\ \cline{2-4}
	& \multicolumn{1}{c|}{Standard}                      & \multicolumn{1}{c|}{2365x1774} & \multicolumn{1}{c|}{1774x2365} & \multicolumn{1}{c|}{}                           & \multicolumn{1}{c|}{}                           & \multicolumn{1}{c|}{}                                                              &                                                               \\ \hline
	Signal                                                       & \multicolumn{1}{c|}{General}                       & \multicolumn{1}{c|}{1536x1152} & \multicolumn{1}{c|}{1152x1536} & \multicolumn{2}{c|}{Indistinguishable}                                                            & \multicolumn{1}{c|}{4096x2304}                                                     & 2304x4096                                                     \\ \hline
	Wickr Me                                                     & \multicolumn{7}{c|}{Indistinguishable}                                                                                                                                                                                                                                                                                                                                        \\ \hline
	\end{tabular}
\end{table*}

\subsection{Image change when sending messenger}
Among the media information of the image, only the resolution of the image transmitted through the messenger can be tracked through footprinting.
The reason is that when comparing Android image and iOS image media information, the extension and format are the same, unlike video media.
Thus, the only way to differentiate them is by resolution. 
For each messenger platform, we conducted experiments on all options such as normal quality, high definition, and original. 
As a result of the experiment, there are many platforms that can be tracked through footprinting using a specific resolution for each instant messenger. 
However, there are several platforms where the original image is transmitted without changing the resolution, or the resolution is changed irregularly, making it impossible to track the footprint.

In the main text, we exluded the results of images with irregularly changed resolution or transmitted as a general original image. 
There are also cases where the error range of the image resolution result is about ±10, but these are very rare.
Table~\ref{tbl6} summarizes the image resolution changes for the 13 messenger platforms that were the subject of the experiment.\\

\textbf{4.2.1. KakaoTalk} \\
KakaoTalk offers three image quality options when transmitting images: normal, high quality, and original. 
Through our experiments, we found that normal and high-quality options, except for the original, can be traced through footprinting using a specific resolution. 
However, the original option has the same resolution as the normal original image, making it impossible to track through the footprint. 
Furthermore, since the resolution of media transmitted from iOS and Android (4032x3024) is the same, it is impossible to distinguish between the operating systems. \\

\textbf{4.2.2. Facebook Messenger} \\
In Facebook Messenger, there is no option to select image quality, thus only basic image quality is available. 
For both iOS and Android (4032x3024), the resolution changes irregularly or is the same as the normal original image resolution, making it impossible to track through footprints. 
However, since Android (5312x2988) uses a specific resolution, it can be tracked by footprint. 
The resolution result has two outcomes, which is different from other messengers. 
Although the result depends on the process of media dissemination, if the same result as shown in Table~\ref{tbl7} is obtained, it can be identified as media transmitted through Facebook Messenger. \\ 

\textbf{4.2.3. Facebook} \\
Facebook does not have a quality option, so only the default image quality exists. In iOS, tracking through footprints is possible using a specific resolution. However, in Android, the resolution changes irregularly, making tracking through footprints impossible. From the resolution result, it can be seen that the normal resolution of KakaoTalk and the default resolution of Facebook are the same. When the same resolution appears, it is possible to distinguish it by a specific platform through an additional file size. The file size of KakaoTalk general definition media is about 100KB, and the file size of Facebook basic definition media is about 50KB. The error range of the two files is about ±10KB. \\

\textbf{4.2.4. Instagram} \\
Instagram does not have a quality option, so only the default image quality exists. Additionally, both iOS and Android have irregular resolutions, making tracking through footprints impossible. \\

\textbf{4.2.5. WhatsApp} \\
WhatsApp does not have a quality option, so only the default image quality exists. Both iOS and Android use specific resolutions to enable tracking through footprints. Additionally, the resolution of the media transferred from iOS and Android (4032x3024) is the same, making it impossible to distinguish the operating system.\\

\textbf{4.2.6. WeChat} \\
WeChat allows to choose between two options, normal quality and full quality to transmit. 
The general image quality can be traced through footprinting using a specific resolution. However, the overall image quality cannot be distinguished because it is the same as the general original image resolution, and tracking through footprinting is impossible. 
Through the resolution result analysis, it can be seen that the WeChat normal resolution and KakaoTalk high-definition resolution produce the same result.
As described above, when the same resolution is obtained, it is possible to distinguish a specific platform through an additional file size. 

The file size of KakaoTalk high-definition media is about 500KB, and the file size of WeChat normal-definition media is about 200KB. The error range of the two files is about ±100KB.
There are cases where the results may differ depending on the dissemination process, but it is very rare. \\

\textbf{4.2.7. Telegram} \\
Telegram can be transmitted by selecting two options: normal quality and original. 
General image quality can be tracked through footprinting using a specific resolution. 
However, since the original option is the same as the normal original image resolution and cannot be distinguished, it is impossible to trace through the footprint. 
Also, since the media transferred from iOS and Android have the same resolution, it is impossible to distinguish between operating systems. \\

\textbf{4.2.8. Skype} \\
Skype has no picture quality options, thus only basic picture quality exists. 
Both iOS and Android use specific resolutions to enable tracking through footprints. \\

\textbf{4.2.9. Discord} \\
In Discord, when downloading an image, it can be chosen between the preview option and the original option. In the preview option, both iOS and Android change resolution irregularly, thus tracking through footprints is impossible. And since the original option produces the same result as the normal original image resolution, tracking through both options is impossible through footprints. \\

\textbf{4.2.10. NateOn} \\
When sending an image in NateOn, iOS users can choose between two options: normal image quality and original image quality, while Android users can select from three options: minimum quality, standard quality, and original quality. For both iOS and Android with a resolution of 4032x3024, the resolution changes irregularly or is the same as the normal original image resolution, making tracking through footprints impossible. However, Android with a resolution of 5312x2988 uses a specific resolution, allowing it to be tracked through footprints. \\

\textbf{4.2.11. LINE} \\
In LINE, it is possible to select from three options: normal quality, standard quality, and original quality. 
Since both iOS and Android use specific resolutions for normal and standard quality, tracking through footprint is possible. 
However, since the original image quality is the same as the general original image resolution, it is impossible to track through footprinting. \\

\textbf{4.2.12. Signal} \\
Signal only has basic image quality as it does not provide quality options. Both iOS and Android allow tracking through footprints using specific resolutions (5312x2988). However, tracking through footprint is impossible in Android (4032x3024) as it's the same as the original image resolution. \\

\textbf{4.2.13. Wickr Me} \\
In Wickr Me, only basic image quality exists as there is no option for image quality. Additionally, tracking through footprints is impossible as both iOS and Android have the same resolution as the normal original image. \\

\begin{table*}
\caption{iOS video changes in instant messengers}\label{tbl7}
\resizebox{\textwidth}{!}{
	\begin{tabular}{|c|c|c|c|c|c|c|c|c|c|}
	\hline
	\begin{tabular}[c]{@{}c@{}}Application \\ name\end{tabular}  & \begin{tabular}[c]{@{}c@{}}Video \\ Quality\end{tabular} & Extension            & Format Profile                        & Codec ID                                    & \begin{tabular}[c]{@{}c@{}}Video Format \\ Profile\end{tabular} & Width                                                     & Length                                                     & Encoder                        & \begin{tabular}[c]{@{}c@{}}Additional \\ Information\end{tabular} \\ \hline
	\multirow{4}{*}{KakaoTalk}                                   & \multirow{2}{*}{General}                                 & \multirow{2}{*}{mp4} & Base Media                            & isom (isom/iso2/avc1/mp41)                  & \multirow{2}{*}{Baseline@L4.1}                                  & \multirow{2}{*}{720x404}                                  & \multirow{2}{*}{404x720}                                   & Lavf57.56.101                  &                                                                   \\ \cline{4-5} \cline{9-10} 
	&                                                          &                      & Base Media Version 2                  & mp42 (isom/mp41/mp42)                       &                                                                 &                                                           &                                                            &                                &                                                                   \\ \cline{2-10} 
	& \multirow{2}{*}{High}                                    & \multirow{2}{*}{mp4} & Base Media                            & isom (isom/iso2/mp41)                       & Main@L4@Main                                                    & \multirow{2}{*}{1920x1080}                                & \multirow{2}{*}{1920x1080}                                 & Lavf57.83.100                  &                                                                   \\ \cline{4-6} \cline{9-10} 
	&                                                          &                      & Base Media Version 2                  & mp42 (isom/mp42)                            & Baseline@L4                                                     &                                                           &                                                            &                                &                                                                   \\ \hline
	\begin{tabular}[c]{@{}c@{}}Facebook\\ Messenger\end{tabular} & Default                                                  & mp4, MOV             & \multirow{2}{*}{Base Media}           & \multirow{2}{*}{isom (isom/iso2/avc1/mp41)} & \multirow{2}{*}{Main@L3.1}                                      & \multirow{2}{*}{1280x720}                                 & \multirow{2}{*}{720x1280}                                  & \multirow{2}{*}{Lavf58.20.100} &                                                                   \\ \cline{1-3} \cline{10-10} 
	Facebook                                                     & Default                                                  & mp4                  &                                       &                                             &                                                                 &                                                           &                                                            &                                & Movie name                                                        \\ \hline
	Instagram                                                    & Default                                                  & mp4, MOV             & Base Media                            & isom (isom/iso2/avc1/mp41)                  & Baseline@L3                                                     & \begin{tabular}[c]{@{}c@{}}480x270\\ 640x360\end{tabular} & \begin{tabular}[c]{@{}c@{}}480x854\\ 640x1138\end{tabular} & Lavf58.20.100                  &                                                                   \\ \hline
	WhatsApp                                                     & Default                                                  & mp4                  & Base Media Version 2                  & mp42 (mp42/isom)                            & Baseline@L3.1                                                   & 848x480                                                   & 848x480                                                    &                                &                                                                   \\ \hline
	WeChat                                                       & Default                                                  & mp4                  & Base Media Version 2                  & mp42 (isom/mp41/mp42)                       & High@L3.1                                                       & 960x544                                                   & 544x960                                                    &                                & Movie More                                                        \\ \hline
	\multirow{5}{*}{Telegram}                                    & 1080p                                                    & \multirow{5}{*}{MOV} & \multirow{5}{*}{Base Media Version 2} & \multirow{5}{*}{mp42 (isom/mp41/mp42)}      & High@L4                                                         & 1920x1072                                                 & 1072x1920                                                  &                                &                                                                   \\ \cline{2-2} \cline{6-10} 
	& 720p                                                     &                      &                                       &                                             & High@L3.1                                                       & 1280x720                                                  & 720x1280                                                   &                                &                                                                   \\ \cline{2-2} \cline{6-10} 
	& 480p                                                     &                      &                                       &                                             & High@L3.1                                                       & 848x464                                                   & 464x848                                                    &                                &                                                                   \\ \cline{2-2} \cline{6-10} 
	& 360p                                                     &                      &                                       &                                             & High@L3                                                         & 640x352                                                   & 352x640                                                    &                                &                                                                   \\ \cline{2-2} \cline{6-10} 
	& 240p                                                     &                      &                                       &                                             & High@L2.1                                                       & 480x256                                                   & 256x480                                                    &                                &                                                                   \\ \hline
	Skype                                                        & Default                                                  & MOV                  & Base Media Version 2                  & mp42 (isom/mp41/mp42)                       & Main@L3.1                                                       & 1280x720                                                  & 720x1280                                                   &                                &                                                                   \\ \hline
	Discord                                                      & Default                                                  & MOV                  & QuickTime                             & qt                                          & Main@L3.1                                                       & 960x540                                                   & 960x540                                                    &                                &                                                                   \\ \hline
	NateOn                                                       & Default                                                  & mp4                  & Base Media Version 2                  & mp42 (isom/mp41/mp42)                       & High@L4                                                         & 1920x1080                                                 & 1920x1080                                                  &                                & Recorded date                                                     \\ \hline
	LINE                                                         & Default                                                  & mp4                  & Base Media Version 2                  & mp42 (isom/mp41/mp42)                       & High@L4                                                         & 960x540                                                   & 960x540                                                    &                                & Movie name                                                        \\ \hline
	Signal                                                       & Default                                                  & mp4                  & Base Media Version 2                  & mp42 (isom/mp41/mp42)                       & Main@L3                                                         & 568x320                                                   & 568x320                                                    &                                &                                                                   \\ \hline
	Wickr Me                                                     & Default                                                  & MOV                  & QuickTime                             & qt                                          & Main@L3                                                         & 568x320                                                   & 568x320                                                    &                                &                                                                   \\ \hline
	\end{tabular}
}
\end{table*}

\begin{table*}
\caption{Android video changes in instant messengers}\label{tbl8}	
\resizebox{\textwidth}{!}{
	\begin{tabular}{|c|ccccccccc|}
	\hline
		\begin{tabular}[c]{@{}c@{}}Application \\ name\end{tabular}  & \multicolumn{1}{c|}{\begin{tabular}[c]{@{}c@{}}Video \\ Quality\end{tabular}}         & \multicolumn{1}{c|}{Extension}            & \multicolumn{1}{c|}{Format Profile}              & \multicolumn{1}{c|}{Codec ID}                                    & \multicolumn{1}{c|}{\begin{tabular}[c]{@{}c@{}}Video Format \\ Profile\end{tabular}} & \multicolumn{1}{c|}{Width}                                                     & \multicolumn{1}{c|}{Length}                                                     & \multicolumn{1}{c|}{Encoder}           & \begin{tabular}[c]{@{}c@{}}Additional \\ Information\end{tabular} \\ \hline
		\multirow{4}{*}{KakaoTalk}                                   & \multicolumn{1}{c|}{\multirow{2}{*}{General}}                                         & \multicolumn{1}{c|}{\multirow{2}{*}{mp4}} & \multicolumn{1}{c|}{Base Media}                  & \multicolumn{1}{c|}{isom (isom/iso2/avc1/mp41)}                  & \multicolumn{1}{c|}{Baseline@L3}                                                     & \multicolumn{1}{c|}{\multirow{2}{*}{852x480}}                                  & \multicolumn{1}{c|}{\multirow{2}{*}{852x480}}                                   & \multicolumn{1}{c|}{Lavf57.56.101}     &                                                                   \\ \cline{4-6} \cline{9-10} 
		& \multicolumn{1}{c|}{}                                                                 & \multicolumn{1}{c|}{}                     & \multicolumn{1}{c|}{Base Media Version 2}        & \multicolumn{1}{c|}{mp42 (isom/mp41/mp42)}                       & \multicolumn{1}{c|}{Baseline@L3.1}                                                   & \multicolumn{1}{c|}{}                                                          & \multicolumn{1}{c|}{}                                                           & \multicolumn{1}{c|}{}                  &                                                                   \\ \cline{2-10} 
		& \multicolumn{1}{c|}{\multirow{2}{*}{High}}                                            & \multicolumn{1}{c|}{\multirow{2}{*}{mp4}} & \multicolumn{1}{c|}{Base Media}                  & \multicolumn{1}{c|}{isom (isom/iso2/mp41)}                       & \multicolumn{1}{c|}{Main@L4@Main}                                                    & \multicolumn{1}{c|}{\multirow{2}{*}{1920x1080}}                                & \multicolumn{1}{c|}{\multirow{2}{*}{1920x1080}}                                 & \multicolumn{1}{c|}{Lavf57.83.100}     &                                                                   \\ \cline{4-6} \cline{9-10} 
		& \multicolumn{1}{c|}{}                                                                 & \multicolumn{1}{c|}{}                     & \multicolumn{1}{c|}{Base Media Version 2}        & \multicolumn{1}{c|}{mp42 (isom/mp42)}                            & \multicolumn{1}{c|}{Baseline@L4}                                                     & \multicolumn{1}{c|}{}                                                          & \multicolumn{1}{c|}{}                                                           & \multicolumn{1}{c|}{}                  &                                                                   \\ \hline
		\begin{tabular}[c]{@{}c@{}}Facebook\\ Messenger\end{tabular} & \multicolumn{1}{c|}{Default}                                                          & \multicolumn{1}{c|}{mp4}                  & \multicolumn{1}{c|}{\multirow{2}{*}{Base Media}} & \multicolumn{1}{c|}{\multirow{2}{*}{isom (isom/iso2/avc1/mp41)}} & \multicolumn{1}{c|}{\begin{tabular}[c]{@{}c@{}}Main@L3\\ Main@L4\end{tabular}}       & \multicolumn{2}{c|}{Resolution irregularity}                                                                                                                     & \multicolumn{1}{c|}{Lavf58.20.100}     &                                                                   \\ \cline{1-3} \cline{6-10} 
		Facebook                                                     & \multicolumn{1}{c|}{Default}                                                          & \multicolumn{1}{c|}{mp4}                  & \multicolumn{1}{c|}{}                            & \multicolumn{1}{c|}{}                                            & \multicolumn{1}{c|}{Baseline@L3}                                                     & \multicolumn{1}{c|}{400x224}                                                   & \multicolumn{1}{c|}{224x400}                                                    & \multicolumn{1}{c|}{Lavf56.40.101}     & Movie name                                                        \\ \hline
		Instagram                                                    & \multicolumn{1}{c|}{Default}                                                          & \multicolumn{1}{c|}{mp4}                  & \multicolumn{1}{c|}{Base Media}                  & \multicolumn{1}{c|}{isom (isom/iso2/avc1/mp41)}                  & \multicolumn{1}{c|}{Baseline@L3}                                                     & \multicolumn{1}{c|}{\begin{tabular}[c]{@{}c@{}}480x270\\ 640x360\end{tabular}} & \multicolumn{1}{c|}{\begin{tabular}[c]{@{}c@{}}480x852\\ 640x1136\end{tabular}} & \multicolumn{1}{c|}{Lavf58.20.100}     &                                                                   \\ \hline
		WhatsApp                                                     & \multicolumn{9}{c|}{Indistinguishable}                                                                                                                                                                                                                                                                                                                                                                                                                                                                                                                                                                                         \\ \hline
		WeChat                                                       & \multicolumn{1}{c|}{Default}                                                          & \multicolumn{1}{c|}{mp4}                  & \multicolumn{1}{c|}{Base Media Version 2}        & \multicolumn{1}{c|}{mp42 (isom/mp41/mp42)}                       & \multicolumn{1}{c|}{High@L3.1}                                                       & \multicolumn{1}{c|}{960x544}                                                   & \multicolumn{1}{c|}{544x960}                                                    & \multicolumn{1}{c|}{}                  & Copyright                                                         \\ \hline
		\multirow{5}{*}{Telegram}                                    & \multicolumn{1}{c|}{High}                                                             & \multicolumn{1}{c|}{\multirow{5}{*}{mp4}} & \multicolumn{1}{c|}{\multirow{5}{*}{Base Media}} & \multicolumn{1}{c|}{\multirow{5}{*}{isom (isom/iso2/avc1/mp41)}} & \multicolumn{1}{c|}{\begin{tabular}[c]{@{}c@{}}High@L4\\ Baseline@L4\end{tabular}}   & \multicolumn{1}{c|}{1920x1080}                                                 & \multicolumn{1}{c|}{1080x1920}                                                  & \multicolumn{1}{c|}{}                  &                                                                   \\ \cline{2-2} \cline{6-10} 
		& \multicolumn{1}{c|}{Medium}                                                           & \multicolumn{1}{c|}{}                     & \multicolumn{1}{c|}{}                            & \multicolumn{1}{c|}{}                                            & \multicolumn{1}{c|}{Baseline@L3.1}                                                   & \multicolumn{1}{c|}{1280x720}                                                  & \multicolumn{1}{c|}{720x1280}                                                   & \multicolumn{1}{c|}{}                  &                                                                   \\ \cline{2-2} \cline{6-10} 
		& \multicolumn{1}{c|}{\begin{tabular}[c]{@{}c@{}}Between\\ medium and low\end{tabular}} & \multicolumn{1}{c|}{}                     & \multicolumn{1}{c|}{}                            & \multicolumn{1}{c|}{}                                            & \multicolumn{1}{c|}{Baseline@L3.1}                                                   & \multicolumn{1}{c|}{854x480}                                                   & \multicolumn{1}{c|}{480x854}                                                    & \multicolumn{1}{c|}{}                  &                                                                   \\ \cline{2-2} \cline{6-10} 
		& \multicolumn{1}{c|}{\multirow{2}{*}{Low}}                                             & \multicolumn{1}{c|}{}                     & \multicolumn{1}{c|}{}                            & \multicolumn{1}{c|}{}                                            & \multicolumn{1}{c|}{\multirow{2}{*}{Baseline@L2.1}}                                  & \multicolumn{1}{c|}{\multirow{2}{*}{480x270}}                                  & \multicolumn{1}{c|}{\multirow{2}{*}{270x480}}                                   & \multicolumn{1}{c|}{\multirow{2}{*}{}} & \multirow{2}{*}{}                                                 \\
		& \multicolumn{1}{c|}{}                                                                 & \multicolumn{1}{c|}{}                     & \multicolumn{1}{c|}{}                            & \multicolumn{1}{c|}{}                                            & \multicolumn{1}{c|}{}                                                                & \multicolumn{1}{c|}{}                                                          & \multicolumn{1}{c|}{}                                                           & \multicolumn{1}{c|}{}                  &                                                                   \\ \hline
		Skype                                                        & \multicolumn{1}{c|}{Default}                                                          & \multicolumn{1}{c|}{mp4}                  & \multicolumn{1}{c|}{Base Media Version 2}        & \multicolumn{1}{c|}{mp42 (isom/mp42)}                            & \multicolumn{1}{c|}{Baseline@L3.1}                                                   & \multicolumn{1}{c|}{1280x720}                                                  & \multicolumn{1}{c|}{720x1280}                                                   & \multicolumn{1}{c|}{}                  &                                                                   \\ \hline
		Discord                                                      & \multicolumn{9}{c|}{Indistinguishable}                                                                                                                                                                                                                                                                                                                                                                                                                                                                                                                                                                                         \\ \hline
		NateOn                                                       & \multicolumn{9}{c|}{Indistinguishable}                                                                                                                                                                                                                                                                                                                                                                                                                                                                                                                                                                                         \\ \hline
		LINE                                                         & \multicolumn{1}{c|}{Default}                                                          & \multicolumn{1}{c|}{mp4}                  & \multicolumn{1}{c|}{Base Media Version 2}        & \multicolumn{1}{c|}{mp42 (isom/mp42)}                            & \multicolumn{1}{c|}{High@L3.1}                                                       & \multicolumn{1}{c|}{960x540}                                                   & \multicolumn{1}{c|}{540x960}                                                    & \multicolumn{1}{c|}{}                  &                                                                   \\ \hline
		Signal                                                       & \multicolumn{1}{c|}{Default}                                                          & \multicolumn{1}{c|}{mp4}                  & \multicolumn{1}{c|}{Base Media Version 2}        & \multicolumn{1}{c|}{mp42 (isom/mp42)}                            & \multicolumn{1}{c|}{Baseline@L3.1}                                                   & \multicolumn{1}{c|}{1280x720}                                                  & \multicolumn{1}{c|}{720x1280}                                                   & \multicolumn{1}{c|}{}                  &                                                                   \\ \hline
		Wickr Me                                                     & \multicolumn{9}{c|}{Indistinguishable}                                                                                                                                                                                                                                                                                                                                                                                                                                                                                                                                                                                         \\ \hline
	\end{tabular}
}
\end{table*}

\subsection{Video change when sending messenger}
When transmitting a video through a messenger, we confirmed that it is possible to track it through footprinting by analyzing various media information such as extension, format profile, codec ID, video format profile, resolution, encoder, and specific information. 
Most of the file sizes were small, and the resolution changed during the encoding process on the messenger platform.
Some messenger platforms can provide encoder and specific information for each platform.
By synthesizing specific resolution, encoder information, media information, and others for each instant messenger, it is possible to distinguish artifacts of a specific messenger platform and track them through footprinting.
However, we also found a few platforms where media information was consistent with the original video, making tracking through footprinting impossible.
We analyzed the video using the default resolution of 1920x1080 for both iOS and Android, excluding results that were the same as the original video in the main text.
Table~\ref{tbl7} and Table~\ref{tbl8} summarize the changes in video for the 13 instant messenger platforms that were tested. \\

\textbf{4.3.1. KakaoTalk} \\
KakaoTalk can be transmitted by selecting between two options: normal quality and high quality.
Although the results may differ depending on the dissemination process, the disseminated media always has one of the two results. 
If an encoder is used, different encoder information can be checked for each image quality, and a specific encoder unique to KakaoTalk has been confirmed. 
Once encoded files are redistributed within the messenger, there is little change in the files. 
If an encoder is used, KakaoTalk's specific encoder, `Lavf57.56.101' for normal quality and `Lavf57.83.100' for high quality, can be confirmed. When distributing through KakaoTalk, artifacts such as resolution and other media information can be aggregated and tracked through footprinting. \\

\textbf{4.3.2. Facebook Messenger} \\
In Facebook Messenger, there is no video quality option, thus only basic video quality exists. 
Facebook Messenger has `Lavf58.20.100' information, which is an encoder common to Facebook and Instagram. 
In addition, iOS changes resolution regularly, while Android changes resolution irregularly. 
When distributing through Facebook Messenger, Android cannot specify the resolution.
However, it can be specified by combining other media information, and tracking through footprint is possible. 
Then, iOS aggregates artifacts such as resolution and other media information, and based on the results, tracking through footprint is possible. \\

\textbf{4.3.3. Facebook} \\
Facebook does not have video quality options, thus only default image quality exists. 
As with KakaoTalk described above, the results vary depending on the dissemination process, however distributed videos always produce one of two results.
Facebook has the same encoder information as Facebook Messenger and Instagram, and Facebook's specific information such as `Movie name' leaves information related to the post number. 
When distributing via Facebook, artifacts such as resolution and other media information can be aggregated and traced through footprint based on the results. \\

\textbf{4.3.4. Instagram} \\
Instagram does not have video quality options, thus only default image quality exists. 
Instagram has the same encoder information as Facebook Messenger or Facebook. 
As a result, Facebook Messenger, Facebook, and Instagram have the same encoder information but can be classified as specific platforms by comprehensively checking media information such as codec ID and video format profile. 
When distributing through Instagram, artifacts such as resolution and other media information can be aggregated and tracked through footprinting. \\

\textbf{4.3.5. WhatsApp} \\
WhatsApp does not have video quality options, thus only default quality exists.
iOS can be identified as a WhatsApp platform, and therefore tracked through footprints.
However, since Android has similar media information to general original video, it is impossible to distinguish it as a specific platform, and therefore impossible to track it through footprints. 
When distributing via WhatsApp, only iOS can aggregate artifacts such as resolution and other media information and track them through footprints. \\

\textbf{4.3.6. WeChat} \\
WeChat does not have video quality options, thus only basic image quality exists.
Both iOS and Android versions of WeChat have specific information unique to WeChat that can be checked.
In iOS, information related to the WeChat version called ``Movie More'' remains, and in Android, information related to copyright called ``Copyright'' remains.
When distributing through WeChat, artifacts such as resolution and other media information can be aggregated and tracked through footprinting. \\

\textbf{4.3.7. Telegram} \\
For Telegram, there are five video quality options for iOS: 1080p, 720p, 480p, 360p, and 240p, and four video quality options for Android: high quality, medium quality, medium-low quality, and low quality.
In Telegram, only high-definition horizontal videos have a video format profile of High on Android, while the video format profile for the rest of the high-definition vertical videos and the medium-, low-mid, and low-quality videos is the baseline.
On Android, high-definition vertical videos have a video format profile of the baseline.
As a result, Telegram has different resolutions and profiles for each video quality, and the distributed video quality can be checked.
When disseminating videos through Telegram, artifacts such as resolution and other media information can be aggregated and traced through footprinting. \\

\textbf{4.3.8. Skype} \\
Skype does not have video quality options, thus only basic quality exists.
Skype can be identified as specific platforms for both iOS and Android. Therefore, when distributing through Skype, artifacts such as resolution and other media information can be aggregated and tracked through footprinting. \\

\textbf{4.3.9. Discord} \\
Discord does not offer video quality options, thus only basic quality is available.
It is possible to classify iOS as a specific platform, but Android transmits the original video without any modifications, making tracking through footprints impossible.
Therefore, when distributing through Discord, only iOS can aggregate artifacts such as resolution and other media information for tracking through footprinting. \\

\textbf{4.3.10. NateOn} \\
In NateOn, there are two video quality options: normal video quality and original video quality in iOS, and only basic video quality without options exists in Android. 
Both options can be distinguished by the specific platform in iOS, but in Android, the original video is transmitted as it is and cannot be distinguished, and tracking through footprinting is also impossible.
NateOn's iOS-specific information will leave information related to the date the video was recorded called `Recorded date'. 
When distributing through NateOn, only iOS can aggregate artifacts such as resolution and other media information and track them through footprinting.\\

\textbf{4.3.11. LINE} \\
In LINE, there are no video quality options, thus only basic video quality exists. iOS can be classified as a specific platform regardless of the dissemination process, but for Android, it depends on whether it can be distinguished or not based on the dissemination process. In this text, we only consider the case where the distinction is possible for Android. LINE videos on iOS contain specific information called `Movie name'. When distributing through LINE, Android media information can be tracked through footprinting only if it can be identified by the LINE platform. Additionally, iOS can aggregate artifacts such as resolution and other media information to enable tracking through footprinting. \\

\textbf{4.3.12. Signal} \\
Signal does not offer video quality options, therefore only basic video quality is available. Additionally, both iOS and Android platforms can be identified specifically when using Signal. When distributing media through Signal, information such as resolution and other artifacts can be aggregated and tracked through footprinting. \\

\textbf{4.3.13. Wickr Me} \\
In Wickr Me, there are no video quality options, thus only basic video quality exists. iOS can be identified as a specific platform, thus tracking through footprints is possible. However, since Android transmits the original video as it is and cannot distinguish it, tracking through footprints is also impossible. When distributing through Wickr Me, only iOS can aggregate artifacts such as resolution and other media information and track them through footprints. \\

\begin{table*}
	\caption{Video media changes from messenger to messenger (KakaoTalk / iOS)}\label{tbl9}
	\resizebox{\textwidth}{!}{%
		\begin{tabular}{|c|cccccccc|}
			\hline
			\begin{tabular}[c]{@{}c@{}}Application \\ name\end{tabular}                   & \multicolumn{1}{c|}{Extension}            & \multicolumn{1}{c|}{Format Profile}                        & \multicolumn{1}{c|}{Codec ID}                               & \multicolumn{1}{c|}{\begin{tabular}[c]{@{}c@{}}Video Format \\ Profile\end{tabular}} & \multicolumn{1}{c|}{Width}                    & \multicolumn{1}{c|}{Length}                   & \multicolumn{1}{c|}{Encoder}       & \begin{tabular}[c]{@{}c@{}}Additional \\ Information\end{tabular} \\ \hline
			\multirow{2}{*}{\begin{tabular}[c]{@{}c@{}}Facebook\\ Messenger\end{tabular}} & \multicolumn{1}{c|}{\multirow{3}{*}{mp4}} & \multicolumn{1}{c|}{Base Media}                            & \multicolumn{1}{c|}{isom (isom/iso2/avc1/mp41)}             & \multicolumn{1}{c|}{Main@L3}                                                         & \multicolumn{1}{c|}{720x404}                  & \multicolumn{1}{c|}{404x720}                  & \multicolumn{1}{c|}{Lavf58.20.100} &                                                                   \\ \cline{3-9} 
			& \multicolumn{1}{c|}{}                     & \multicolumn{1}{c|}{Base Media Version 2}                  & \multicolumn{1}{c|}{mp42 (isom/mp41/mp42)}                  & \multicolumn{1}{c|}{Baseline@L4.1}                                                   & \multicolumn{1}{c|}{720x404}                  & \multicolumn{1}{c|}{720x404}                  & \multicolumn{1}{c|}{}              &                                                                   \\ \cline{1-1} \cline{3-9} 
			Facebook                                                                      & \multicolumn{1}{c|}{}                     & \multicolumn{1}{c|}{Base Media}                            & \multicolumn{1}{c|}{isom (isom/iso2/avc1/mp41)}             & \multicolumn{1}{c|}{Main@L3}                                                         & \multicolumn{1}{c|}{640x358}                  & \multicolumn{1}{c|}{358x640}                  & \multicolumn{1}{c|}{Lavf58.20.100} & Movie name                                                        \\ \hline
			Instagram                                                                     & \multicolumn{8}{c|}{Indistinguishable}                                                                                                                                                                                                                                                                                                                                                                                                                               \\ \hline
			WhatsApp                                                                      & \multicolumn{1}{c|}{\multirow{2}{*}{mp4}} & \multicolumn{1}{c|}{\multirow{2}{*}{Base Media Version 2}} & \multicolumn{1}{c|}{mp42 (mp42/isom)}                       & \multicolumn{1}{c|}{Baseline@L3}                                                     & \multicolumn{1}{c|}{\multirow{2}{*}{720x404}} & \multicolumn{1}{c|}{720x404}                  & \multicolumn{1}{c|}{}              &                                                                   \\ \cline{1-1} \cline{4-5} \cline{7-9} 
			WeChat                                                                        & \multicolumn{1}{c|}{}                     & \multicolumn{1}{c|}{}                                      & \multicolumn{1}{c|}{mp42 (isom/mp41/mp42)}                  & \multicolumn{1}{c|}{High@L3}                                                         & \multicolumn{1}{c|}{}                         & \multicolumn{1}{c|}{404x720}                  & \multicolumn{1}{c|}{}              & Movie More                                                        \\ \hline
			Telegram                                                                      & \multicolumn{8}{c|}{Indistinguishable}                                                                                                                                                                                                                                                                                                                                                                                                                               \\ \hline
			Skype                                                                         & \multicolumn{1}{c|}{\multirow{6}{*}{mp4}} & \multicolumn{1}{c|}{\multirow{5}{*}{Base Media Version 2}} & \multicolumn{1}{c|}{\multirow{5}{*}{mp42 (isom/mp41/mp42)}} & \multicolumn{1}{c|}{\multirow{3}{*}{Baseline@L4.1}}                                  & \multicolumn{1}{c|}{\multirow{4}{*}{720x404}} & \multicolumn{1}{c|}{\multirow{3}{*}{720x404}} & \multicolumn{1}{c|}{}              &                                                                   \\ \cline{1-1} \cline{8-9} 
			Discord                                                                       & \multicolumn{1}{c|}{}                     & \multicolumn{1}{c|}{}                                      & \multicolumn{1}{c|}{}                                       & \multicolumn{1}{c|}{}                                                                & \multicolumn{1}{c|}{}                         & \multicolumn{1}{c|}{}                         & \multicolumn{1}{c|}{}              &                                                                   \\ \cline{1-1} \cline{8-9} 
			NateOn                                                                        & \multicolumn{1}{c|}{}                     & \multicolumn{1}{c|}{}                                      & \multicolumn{1}{c|}{}                                       & \multicolumn{1}{c|}{}                                                                & \multicolumn{1}{c|}{}                         & \multicolumn{1}{c|}{}                         & \multicolumn{1}{c|}{}              &                                                                   \\ \cline{1-1} \cline{5-5} \cline{7-9} 
			LINE                                                                          & \multicolumn{1}{c|}{}                     & \multicolumn{1}{c|}{}                                      & \multicolumn{1}{c|}{}                                       & \multicolumn{1}{c|}{High@L4}                                                         & \multicolumn{1}{c|}{}                         & \multicolumn{1}{c|}{404x720}                  & \multicolumn{1}{c|}{}              & Line Video                                                        \\ \cline{1-1} \cline{5-9} 
			Signal                                                                        & \multicolumn{1}{c|}{}                     & \multicolumn{1}{c|}{}                                      & \multicolumn{1}{c|}{}                                       & \multicolumn{1}{c|}{Main@L2.1}                                                       & \multicolumn{2}{c|}{\multirow{2}{*}{480x268}}                                                 & \multicolumn{1}{c|}{}              &                                                                   \\ \cline{1-1} \cline{3-5} \cline{8-9} 
			Wickr Me                                                                      & \multicolumn{1}{c|}{}                     & \multicolumn{1}{c|}{QuickTime}                             & \multicolumn{1}{c|}{qt}                                     & \multicolumn{1}{c|}{Main@L2.1}                                                       & \multicolumn{2}{c|}{}                                                                         & \multicolumn{1}{c|}{}              &                                                                   \\ \hline
		\end{tabular}
	}
\end{table*}

\begin{table*}[]
	\caption{Video media changes from messenger to messenger (KakaoTalk / Android)}\label{tbl10}
	\resizebox{\textwidth}{!}{%
		\begin{tabular}{|c|cccccccc|}
			\hline
			\begin{tabular}[c]{@{}c@{}}Application \\ name\end{tabular}  & \multicolumn{1}{c|}{Extension}            & \multicolumn{1}{c|}{Format Profile}                        & \multicolumn{1}{c|}{Codec ID}                                    & \multicolumn{1}{c|}{\begin{tabular}[c]{@{}c@{}}Video Format \\ Profile\end{tabular}} & \multicolumn{1}{c|}{Width}                    & \multicolumn{1}{c|}{Length}                   & \multicolumn{1}{c|}{Encoder}                        & \begin{tabular}[c]{@{}c@{}}Additional \\ Information\end{tabular} \\ \hline
			\begin{tabular}[c]{@{}c@{}}Facebook\\ Messenger\end{tabular} & \multicolumn{1}{c|}{\multirow{2}{*}{mp4}} & \multicolumn{1}{c|}{\multirow{2}{*}{Base Media}}           & \multicolumn{1}{c|}{\multirow{2}{*}{isom (isom/iso2/avc1/mp41)}} & \multicolumn{1}{c|}{\multirow{2}{*}{Main@L3}}                                        & \multicolumn{1}{c|}{\multirow{2}{*}{852x480}} & \multicolumn{1}{c|}{\multirow{2}{*}{480x852}} & \multicolumn{1}{c|}{\multirow{2}{*}{Lavf58.20.100}} &                                                                   \\ \cline{1-1} \cline{9-9} 
			Facebook                                                     & \multicolumn{1}{c|}{}                     & \multicolumn{1}{c|}{}                                      & \multicolumn{1}{c|}{}                                            & \multicolumn{1}{c|}{}                                                                & \multicolumn{1}{c|}{}                         & \multicolumn{1}{c|}{}                         & \multicolumn{1}{c|}{}                               & Movie name                                                        \\ \hline
			Instagram                                                    & \multicolumn{8}{c|}{Indistinguishable}                                                                                                                                                                                                                                                                                                                                                                                                                                                     \\ \hline
			WhatsApp                                                     & \multicolumn{1}{c|}{\multirow{7}{*}{mp4}} & \multicolumn{1}{c|}{\multirow{2}{*}{Base Media Version 2}} & \multicolumn{1}{c|}{\multirow{2}{*}{mp42 (mp42/isom)}}           & \multicolumn{1}{c|}{Baseline@L3.1}                                                   & \multicolumn{1}{c|}{\multirow{7}{*}{852x480}} & \multicolumn{1}{c|}{852x480}                  & \multicolumn{1}{c|}{}                               &                                                                   \\ \cline{1-1} \cline{5-5} \cline{7-9} 
			WeChat                                                       & \multicolumn{1}{c|}{}                     & \multicolumn{1}{c|}{}                                      & \multicolumn{1}{c|}{}                                            & \multicolumn{1}{c|}{High@L3.1}                                                       & \multicolumn{1}{c|}{}                         & \multicolumn{1}{c|}{\multirow{3}{*}{480x852}} & \multicolumn{1}{c|}{}                               &                                                                   \\ \cline{1-1} \cline{3-5} \cline{8-9} 
			Telegram                                                     & \multicolumn{1}{c|}{}                     & \multicolumn{1}{c|}{Base Media}                            & \multicolumn{1}{c|}{isom (isom/iso2/avc1/mp41)}                  & \multicolumn{1}{c|}{\multirow{4}{*}{Baseline@L3.1}}                                  & \multicolumn{1}{c|}{}                         & \multicolumn{1}{c|}{}                         & \multicolumn{1}{c|}{}                               & Copyright                                                         \\ \cline{1-1} \cline{3-4} \cline{8-9} 
			Skype                                                        & \multicolumn{1}{c|}{}                     & \multicolumn{1}{c|}{\multirow{4}{*}{Base Media Version 2}} & \multicolumn{1}{c|}{\multirow{4}{*}{mp42 (isom/mp42)}}           & \multicolumn{1}{c|}{}                                                                & \multicolumn{1}{c|}{}                         & \multicolumn{1}{c|}{}                         & \multicolumn{1}{c|}{}                               &                                                                   \\ \cline{1-1} \cline{7-9} 
			Discord                                                      & \multicolumn{1}{c|}{}                     & \multicolumn{1}{c|}{}                                      & \multicolumn{1}{c|}{}                                            & \multicolumn{1}{c|}{}                                                                & \multicolumn{1}{c|}{}                         & \multicolumn{1}{c|}{\multirow{2}{*}{852x480}} & \multicolumn{1}{c|}{}                               &                                                                   \\ \cline{1-1} \cline{8-9} 
			NateOn                                                       & \multicolumn{1}{c|}{}                     & \multicolumn{1}{c|}{}                                      & \multicolumn{1}{c|}{}                                            & \multicolumn{1}{c|}{}                                                                & \multicolumn{1}{c|}{}                         & \multicolumn{1}{c|}{}                         & \multicolumn{1}{c|}{}                               &                                                                   \\ \cline{1-1} \cline{5-5} \cline{7-9} 
			LINE                                                         & \multicolumn{1}{c|}{}                     & \multicolumn{1}{c|}{}                                      & \multicolumn{1}{c|}{}                                            & \multicolumn{1}{c|}{High@L3.1}                                                       & \multicolumn{1}{c|}{}                         & \multicolumn{1}{c|}{404x720}                  & \multicolumn{1}{c|}{}                               &                                                                   \\ \hline
			Signal                                                       & \multicolumn{8}{c|}{Indistinguishable}                                                                                                                                                                                                                                                                                                                                                                                                                                                     \\ \hline
			Wickr Me                                                     & \multicolumn{1}{c|}{mp4}                  & \multicolumn{1}{c|}{Base Media Version 2}                  & \multicolumn{1}{c|}{mp42 (isom/mp42)}                            & \multicolumn{1}{c|}{Baseline@L3.1}                                                   & \multicolumn{1}{c|}{852x480}                  & \multicolumn{1}{c|}{852x480}                  & \multicolumn{1}{c|}{}                               &                                                                   \\ \hline
		\end{tabular}
	}
\end{table*}

\subsection{Video media switching when sending multiple messages via messenger}
In the previous section, we identified the artifacts of an instant messenger by tracking footprints when video media was distributed once. In this section, we analyze each messenger artifact when it's transmitted from one messenger to another, focusing on normal and basic image quality only. We chose KakaoTalk and Facebook Messenger, two popular messengers in the world, as the N-th distribution sites, and the rest of the messengers as the N+1 distribution sites. Based on the study's results, we investigate whether it's possible to track the N-th and N+1 distribution messengers through footprints.\\

\textbf{4.4.1. N-th messenger KakaoTalk} \\
This section focuses on the changes that occur when KakaoTalk is the N-th messenger, and the remaining messengers are the N+1 messenger. The experiment resulted in three cases. First, it's possible to partially check the media information of both the N-th and N+1st messengers, enabling tracking through footprinting. Second, the media information of the N-th messenger is overwritten with that of the N+1st messenger, making tracking impossible. Finally, there's a case where multiple final media information results are the same, narrowing down the investigation network's scope. The study excluded results that couldn't be traced through footprinting. 
Table~\ref{tbl9} and Table~\ref{tbl10} present the changes in the N-th messenger for KakaoTalk in the iOS and Android environments.
\\

\begin{table*}
\caption{Video media changes from messenger to messenger (Facebook Messenger / iOS)}\label{tbl11}
\resizebox{\textwidth}{!}{%
	\begin{tabular}{|c|cccccccc|}
		\hline
		\begin{tabular}[c]{@{}c@{}}Application \\ name\end{tabular} & \multicolumn{1}{c|}{Extension}            & \multicolumn{1}{c|}{Format Profile}       & \multicolumn{1}{c|}{Codec ID}                   & \multicolumn{1}{c|}{\begin{tabular}[c]{@{}c@{}}Video Format \\ Profile\end{tabular}} & \multicolumn{1}{c|}{Width}    & \multicolumn{1}{c|}{Length}   & \multicolumn{1}{c|}{Encoder}                        & \begin{tabular}[c]{@{}c@{}}Additional \\ Information\end{tabular} \\ \hline
		\multicolumn{1}{|l|}{KakaoTalk}                             & \multicolumn{8}{c|}{Indistinguishable}                                                                                                                                                                                                                                                                                                                                                                                   \\ \hline
		Facebook                                                    & \multicolumn{1}{c|}{mp4}                  & \multicolumn{1}{c|}{Base Media}           & \multicolumn{1}{c|}{isom (isom/iso2/avc1/mp41)} & \multicolumn{1}{c|}{Main@L3.1}                                                       & \multicolumn{1}{c|}{1280x720} & \multicolumn{1}{c|}{720x1280} & \multicolumn{1}{c|}{Lavf58.20.100}                  & Movie name                                                        \\ \hline
		Instagram                                                   & \multicolumn{8}{c|}{Indistinguishable}                                                                                                                                                                                                                                                                                                                                                                                   \\ \hline
		WhatsApp                                                    & \multicolumn{1}{c|}{mp4}                  & \multicolumn{1}{c|}{Base Media Version 2} & \multicolumn{1}{c|}{mp42 (mp42/isom)}           & \multicolumn{1}{c|}{Main@L3.1}                                                       & \multicolumn{1}{c|}{1280x720} & \multicolumn{1}{c|}{720x1280} & \multicolumn{1}{c|}{}                               &                                                                   \\ \hline
		WeChat                                                      & \multicolumn{8}{c|}{Indistinguishable}                                                                                                                                                                                                                                                                                                                                                                                   \\ \hline
		Telegram                                                    & \multicolumn{8}{c|}{Indistinguishable}                                                                                                                                                                                                                                                                                                                                                                                   \\ \hline
		Skype                                                       & \multicolumn{1}{c|}{\multirow{2}{*}{MOV}} & \multicolumn{1}{c|}{Base Media}           & \multicolumn{1}{c|}{isom (isom/iso2/avc1/mp41)} & \multicolumn{1}{c|}{\multirow{3}{*}{Main@L3.1}}                                      & \multicolumn{1}{c|}{1280x720} & \multicolumn{1}{c|}{720x1280} & \multicolumn{1}{c|}{\multirow{2}{*}{Lavf58.20.100}} &                                                                   \\ \cline{1-1} \cline{3-4} \cline{6-7} \cline{9-9} 
		Discord                                                     & \multicolumn{1}{c|}{}                     & \multicolumn{1}{c|}{QuickTime}            & \multicolumn{1}{c|}{qt}                         & \multicolumn{1}{c|}{}                                                                & \multicolumn{1}{c|}{960x540}  & \multicolumn{1}{c|}{540x960}  & \multicolumn{1}{c|}{}                               &                                                                   \\ \cline{1-4} \cline{6-9} 
		NateOn                                                      & \multicolumn{1}{c|}{mp4}                  & \multicolumn{1}{c|}{Base Media Version 2} & \multicolumn{1}{c|}{isom (isom/mp41/mp42)}      & \multicolumn{1}{c|}{}                                                                & \multicolumn{1}{c|}{1280x720} & \multicolumn{1}{c|}{720x1280} & \multicolumn{1}{c|}{}                               &                                                                   \\ \hline
		LINE                                                        & \multicolumn{8}{c|}{Indistinguishable}                                                                                                                                                                                                                                                                                                                                                                                   \\ \hline
		Signal                                                      & \multicolumn{8}{c|}{Indistinguishable}                                                                                                                                                                                                                                                                                                                                                                                   \\ \hline
		Wickr Me                                                    & \multicolumn{1}{c|}{MOV}                  & \multicolumn{1}{c|}{QuickTime}            & \multicolumn{1}{c|}{qt}                         & \multicolumn{1}{c|}{Main@L3}                                                         & \multicolumn{1}{c|}{568x320}  & \multicolumn{1}{c|}{320x568}  & \multicolumn{1}{c|}{Lavf58.20.100}                  &                                                                   \\ \hline
	\end{tabular}
}
\end{table*}

\begin{table*}
\caption{Video media changes from messenger to messenger (Facebook Messenger / Android)}\label{tbl12}
\resizebox{\textwidth}{!}{%
	\begin{tabular}{|c|ccccccc|}
		\hline
		\begin{tabular}[c]{@{}c@{}}Application \\ name\end{tabular} & \multicolumn{1}{c|}{Extension}            & \multicolumn{1}{c|}{Format Profile}              & \multicolumn{1}{c|}{Codec ID}                                    & \multicolumn{1}{c|}{\begin{tabular}[c]{@{}c@{}}Video Format \\ Profile\end{tabular}} & \multicolumn{1}{c|}{Width}                      & \multicolumn{1}{c|}{Length}                     & Encoder                        \\ \hline
		KakaoTalk                                                   & \multicolumn{1}{c|}{\multirow{8}{*}{mp4}} & \multicolumn{1}{c|}{\multirow{8}{*}{Base Media}} & \multicolumn{1}{c|}{\multirow{8}{*}{isom (isom/iso2/avc1/mp41)}} & \multicolumn{1}{c|}{\multirow{8}{*}{Main@L4}}                                        & \multicolumn{1}{c|}{\multirow{8}{*}{1920x1080}} & \multicolumn{1}{c|}{\multirow{8}{*}{1080x1920}} & \multirow{8}{*}{Lavf58.20.100} \\ \cline{1-1}
		Instagram                                                   & \multicolumn{1}{c|}{}                     & \multicolumn{1}{c|}{}                            & \multicolumn{1}{c|}{}                                            & \multicolumn{1}{c|}{}                                                                & \multicolumn{1}{c|}{}                           & \multicolumn{1}{c|}{}                           &                                \\ \cline{1-1}
		WhatsApp                                                    & \multicolumn{1}{c|}{}                     & \multicolumn{1}{c|}{}                            & \multicolumn{1}{c|}{}                                            & \multicolumn{1}{c|}{}                                                                & \multicolumn{1}{c|}{}                           & \multicolumn{1}{c|}{}                           &                                \\ \cline{1-1}
		Telegram                                                    & \multicolumn{1}{c|}{}                     & \multicolumn{1}{c|}{}                            & \multicolumn{1}{c|}{}                                            & \multicolumn{1}{c|}{}                                                                & \multicolumn{1}{c|}{}                           & \multicolumn{1}{c|}{}                           &                                \\ \cline{1-1}
		Discord                                                     & \multicolumn{1}{c|}{}                     & \multicolumn{1}{c|}{}                            & \multicolumn{1}{c|}{}                                            & \multicolumn{1}{c|}{}                                                                & \multicolumn{1}{c|}{}                           & \multicolumn{1}{c|}{}                           &                                \\ \cline{1-1}
		NateOn                                                      & \multicolumn{1}{c|}{}                     & \multicolumn{1}{c|}{}                            & \multicolumn{1}{c|}{}                                            & \multicolumn{1}{c|}{}                                                                & \multicolumn{1}{c|}{}                           & \multicolumn{1}{c|}{}                           &                                \\ \cline{1-1}
		LINE                                                        & \multicolumn{1}{c|}{}                     & \multicolumn{1}{c|}{}                            & \multicolumn{1}{c|}{}                                            & \multicolumn{1}{c|}{}                                                                & \multicolumn{1}{c|}{}                           & \multicolumn{1}{c|}{}                           &                                \\ \cline{1-1}
		Wickr Me                                                    & \multicolumn{1}{c|}{}                     & \multicolumn{1}{c|}{}                            & \multicolumn{1}{c|}{}                                            & \multicolumn{1}{c|}{}                                                                & \multicolumn{1}{c|}{}                           & \multicolumn{1}{c|}{}                           &                                \\ \hline
		Facebook                                                    & \multicolumn{7}{c|}{\multirow{4}{*}{Indistinguishable}}                                                                                                                                                                                                                                                                                                                                     \\ \cline{1-1}
		WeChat                                                      & \multicolumn{7}{c|}{}                                                                                                                                                                                                                                                                                                                                                                       \\ \cline{1-1}
		Skype                                                       & \multicolumn{7}{c|}{}                                                                                                                                                                                                                                                                                                                                                                       \\ \cline{1-1}
		Signal                                                      & \multicolumn{7}{c|}{}                                                                                                                                                                                                                                                                                                                                                                       \\ \hline
	\end{tabular}
}
\end{table*}

\textbf{4.4.2. N-th messenger Facebook Messenger} \\ 
This section analyzes the changes that occur when passing through Facebook Messenger as the N-th messenger, and the remaining messengers as the N+1st messenger. The experiment resulted in three cases, similar to section 4.4.1. First, it's possible to partially check the media information of both the N-th and N+1st messengers, enabling tracking through footprinting. Second, the media information of the N-th messenger is overwritten with that of the N+1st messenger, making tracking impossible. Finally, there's a case where multiple final media information results are the same, narrowing down the investigation network's scope. Unlike the KaKaoTalk case, there are cases where only the investigation network's scope can be narrowed in the Android environment for Facebook Messenger. Some messengers also produced results that couldn't be traced through footprinting.
Table~\ref{tbl11} and Table~\ref{tbl12} present the changes in the N-th messenger for Facebook Messenger in the iOS and Android environments.
\\

\section{Conclusion}
We analyzed the experimental results related to footprint tracking through media information analysis when sending images and video media in instant messengers. The study confirmed that instant messengers work differently on iOS and Android. Furthermore, when a messenger was transferred from one messenger to another, it was possible to check which messengers were passed due to the artifacts of the transmitted messengers remaining. From a digital forensic perspective, we drew the following conclusions:

When an original or similar image exists, the distribution site of the transmitted image can be determined through resolution comparison. Each instant messenger has a different resolution for transmitted images, except for messengers whose resolution is unchanged or irregular. Additionally, while some messengers have the same resolutions on both iOS and Android, others have different resolutions. Thus, it's possible to track the distribution site of the image through footprinting.

In the case where there is an original or a similar video, the distribution of the transmitted video can be checked through media information comparison.
In the video transmitted from instant messengers, different media information remains for each messenger, except for the messenger in which the media information is not changed.
In addition, unlike images, videos exist only when the media information of iOS and Android is different for each messenger.
Therefore, it is possible to estimate the operating system of the distributed smartphone, and it is possible to track the footprint in relation to the distribution site of the video.

When a messenger is transmitted to another messenger, the distribution of the transmitted videos can be confirmed or estimated by checking the media information, and the scope of the investigation network can be narrowed.
We conducted an experiment with KakaoTalk and Facebook Messenger as the N-th messengers, and the remaining messengers as the N+1st messenger.
Finally, we concluded three key findings based on our experiment.

\begin{enumerate}
\item The media information of the N-th messenger and the media information of the N+1st messenger remain partially. Therefore, it is possible to accurately check which messengers have been passed through and track the distribution of the transmitted videos through footprinting.
\item The media information of the N-th messenger is overwritten with the media information of the N+1st messenger. Therefore, it is impossible to confirm exactly which messengers have been passed, and tracking through footprints is not possible.
\item This is a case where the final media information result is the same in many cases. In this case, it is impossible to confirm exactly which messengers have been passed through, however, it is possible to narrow the scope of the investigation network.
\end{enumerate}

We hope that this study will assist digital forensic investigators in swiftly conducting criminal investigations pertaining to media distribution by synthesizing the information presented and tracking the channels through which the media was disseminated.

\end{document}